\documentclass[a4paper,twocolumn,preprintnumbers,citeautoscript,aps,prd,longbibliography,superscriptaddress,floatfix,nofootinbib]{revtex4-2}
\usepackage{amsmath,amsfonts,amssymb,graphics}
\usepackage[normalem]{ulem}
\usepackage{url}
\usepackage{graphicx}
\usepackage{cancel}

\usepackage{mathrsfs}
\usepackage{bbm}
\usepackage{pifont}
\usepackage{braket}
\usepackage{siunitx}
\usepackage{slashed}

\usepackage{mathtools}
\usepackage{xspace}
\usepackage{booktabs}
\usepackage{csquotes}
\usepackage[dvipsnames,svgnames]{xcolor}
\usepackage{tikz}
\usetikzlibrary{positioning,shapes,arrows}
\usetikzlibrary{decorations.pathmorphing}
\usetikzlibrary{decorations.markings}
\usepackage{standalone}

\usepackage{url}
\usepackage[bookmarks=false, breaklinks]{hyperref} 

\definecolor{nicered}{rgb}{0.5,.0,.0}
\definecolor{darkblue}{rgb}{0,.1,.9}
\definecolor{lightblue}{rgb}{0,.1,.6}
\definecolor{applegreen}{rgb}{0.55, 0.71, 0.0}
\definecolor{darkgreen}{rgb}{0.0, 0.2, 0.13}
 
\hypersetup{colorlinks, citecolor=darkgreen ,linkcolor=nicered, urlcolor= lightblue}

\begin{document}
\title{Dynamical origin of neutrino masses and dark matter from a new confining sector}
\author{Maximilian Berbig}
\email{berbig@ific.uv.es}
\affiliation{Departament de Física Teòrica, Universitat de València, 46100 Burjassot, Spain}
\affiliation{Instituto de Física Corpuscular (CSIC-Universitat de València), Parc Científic UV, C/Catedrático
José Beltrán, 2, E-46980 Paterna, Spain}

\author{Juan Herrero-Garcia}
\email{juan.herrero@ific.uv.es}
\affiliation{Departament de Física Teòrica, Universitat de València, 46100 Burjassot, Spain}
\affiliation{Instituto de Física Corpuscular (CSIC-Universitat de València), Parc Científic UV, C/Catedrático
José Beltrán, 2, E-46980 Paterna, Spain}
 
\author{Giacomo Landini}
\email{giacomo.landini@ific.uv.es}
\affiliation{Instituto de Física Corpuscular (CSIC-Universitat de València), Parc Científic UV, C/Catedrático
José Beltrán, 2, E-46980 Paterna, Spain}

\date{\today}

\begin{abstract}
A dynamical mechanism, based on a confining non-abelian dark symmetry, 
which generates Majorana masses for hypercharge-less fermions, is proposed. We apply it to the inverse seesaw scenario, which allows to generate light neutrino masses from the interplay of TeV-scale Pseudo-Dirac mass terms and a small explicit breaking of lepton number. A single generation of vector-like dark quarks, transforming under a $\text{SU}(3)_\text{D}$ gauge symmetry, is coupled to a real singlet scalar, which serves as a portal between the dark quark condensate and three generations of heavy sterile neutrinos. Such a  dark sector and the Standard Model (SM) are kept in thermal equilibrium with each other via sizeable Yukawa couplings to the heavy neutrinos. In this framework the lightest dark baryon, which has spin $3/2$ and is stabilized at the renormalizable level by an accidental dark baryon number symmetry, can account for the observed relic density via thermal freeze-out from annihilations into the lightest dark mesons. These mesons in turn decay to heavy neutrinos, which produce SM final states upon decay.
This model may be probed by next generation neutrino telescopes via neutrino lines produced from dark matter annihilations. 
\end{abstract}

\maketitle

\section{Introduction}

\begin{figure}[t]
 \centering
  \includegraphics[width=0.3\textwidth]{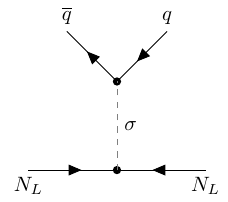}
  \caption{Diagrammatic representation of the origin for the Majorana mass for $\overline{N_L^c} N_L$ from the chiral quark condensate $\braket{\overline{q}q}$. The same diagram also exists for $\overline{N_R^c} N_R$. The mechanism may be used to generate Majorana masses of heavy singlet/triplet fermions in seesaw Type I/III, as well as the $\mu$ term in inverse seesaw.}
  \label{fig:numass}
\end{figure}

The tiny observed mass scale of the active neutrinos is often explained via dynamical mechanisms that avoid the need of considering a small ad-hoc Yukawa coupling by hand. Typically one either suppresses the contribution to $m_\nu$ by the mass scale of a heavy mediator or one introduces  separate scalars with vacuum expectation values (vevs) below the electroweak scale, see for instance Ref.~\cite{Giarnetti:2023xtq}. In this work, we consider the inverse seesaw mechanism \cite{Mohapatra:1986bd,Gonzalez-Garcia:1988okv} (see also Ref.~\cite{Cai:2017jrq}), where one combines both aforementioned ingredients by invoking heavy vector-like fermionic messengers, denoted as $N$ from now on, and introduces a small explicit or spontaneous breaking $\mu$ of lepton number (or B-L). Its main advantage, compared to the usual high-scale seesaw paradigm, is that it may be tested in low-scale experiments. However, note that the mechanism proposed in this work may be also used to generate the high-scale sterile neutrino Majorana mass present in the Type I/III seesaw \cite{Minkowski:1977sc,Yanagida:1979as,Gell-Mann:1979vob,Glashow:1979nm,Yanagida:1980xy, PhysRevLett.44.912, Foot:1988aq}, as well as light sterile Majorana masses (e.g. eV to keV scale). 

Parametrically, in the inverse seesaw scheme the resulting active neutrino mass scales as 
\begin{align}\label{eq:ISS}
		m_\nu\simeq \SI{0.05}{\electronvolt} \cdot y_\nu^2 \left(\frac{\mu}{\SI{1}{\kilo\electronvolt}}\right) \left(\frac{\SI{35}{\tera\electronvolt}}{M_D}\right)^2,
\end{align}
where $M_D$ is the Dirac mass connecting both chiralities of $N$ and $y_\nu$ is  the Yukawa coupling of $N$ to the active neutrinos.
While the original formulation of this mechanism in Ref.~\cite{Mohapatra:1986bd}, which was based on additional singlet fermions added to a supersymmetric $\text{E}_6$ GUT \cite{Witten:1985bz}, assumed that this $\mu$ might arise from Supersymmetry breaking, its relative smallness is often left unexplained. However, models with additional fermion singlets that generate $\mu$ radiatively have been proposed, see Refs.~\cite{Ma:2009gu,Ahriche:2016acx,Coito:2022kif}.

Here we pursue the idea  that such a small breaking of lepton number might arise not from the vev of an elementary scalar but \textit{dynamically} from the formation of a dark quark condensate in a non-abelian dark gauge theory. This idea was first applied  in Ref.~\cite{Thomas:1992hf} to active neutrino masses in a field theory  context  by employing QCD and higher dimensional operators connecting neutrinos to the quark condensate (see Ref.~\cite{Ibanez:2001nd} for a string-theoretic realization and Refs.~\cite{McDonald:1996cs,Davoudiasl:2005ai,Babic:2019zqu,Davoudiasl:2021nfv} for further investigations of the associated phenomenology).
However such an approach requires at least one generation of massless quarks (unless additional model building steps are undertaken \cite{Davoudiasl:2021nfv}), since the quarks are charged under the same symmetry that ensures the absence of the renormalizable neutrino mass terms, which is however heavily disfavored by recent lattice studies \cite{FermilabLattice:2018est,Alexandrou:2020bkd}. 

A compelling alternative is to consider an additional confining gauge group, which only acts on the dark sector, and to use this to generate a small Majorana mass for one chirality of the gauge singlet $N$, which evades the complications accompanying electroweak representations. Additionally the required dark sector can have the right ingredients for a successful dark matter (DM) candidate, which in our case is the lightest dark baryon $\mathcal{B}$. The presence of $N$ with unsuppressed Yukawa couplings to both sectors turns out to be crucial for maintaining thermal equilibrium between them, which allows us to produce the dark matter abundance via thermal freeze-out of annihilations between $\mathcal{B}$  and the lightest dark meson $\mathcal{M}$. References \cite{Hur:2007uz,Kubo:2014ova} used a similar but scale-invariant setup  for generating  electroweak symmetry breaking. When it comes to neutrino masses in scale-invariant frameworks, Ref.~\cite{Hur:2011sv}  generated (at most) TeV-scale right-handed neutrino masses required for realizations of the low scale \cite{Gluza:2002vs,Kersten:2007vk} Type I seesaw \cite{Minkowski:1977sc,Yanagida:1979as,Gell-Mann:1979vob,Glashow:1979nm,  PhysRevLett.44.912} from dynamical chiral symmetry breaking.  The authors of Refs.~\cite{Aoki:2020mlo,Aoki:2021skm} used a similar idea in order to explain a right-handed neutrino mass scale of $\mathcal{O}(10^7\;\text{GeV})$  in the context of the neutrino option \cite{Brivio:2017dfq,Brivio:2018rzm}. Some of the aforementioned works considered dark matter in the form of dark pions produced via e.g. the Higgs-portal, whereas we consider dark baryons with the neutrino portal playing an important role in their thermalization with the SM (for asymmetric DM connected to the SM via the neutrino portal see refs.~\cite{Hall:2019ank,Hall:2019rld,Hall:2021zsk}). Also our framework is not scale invariant and relies on a positive mass squared $m_\sigma^2$ for the singlet scalar.

This article is structured as follows: In Section~\ref{sec:fields} we introduce the field content, in Section~\ref{sec:spec} we specify the particle spectrum and in Section~\ref{sec:cosmo} we discuss the cosmological history. In Section~\ref{sec:signals} we elaborate on possible signatures for indirect detection and in Section~\ref{sec:concl} we summarize our main results.
 
\section{The Model}\label{sec:fields}


\begin{table}[t]
\centering
 \begin{tabular}{|c||c|c|c||c|}
			\hline
			& $\text{SU}(3)_\text{D}$ & $\mathcal{Z}_4$ & $\text{U}(1)_{\text{D}}$  &\text{generations}\\ 
           \hline
			$q_L$ & $\textbf{3}$ &$-i$& $1$&   $1$ \\
			$q_R$ & $\textbf{3}$ & $i$ & $1$ &  $1$ \\
			$N_{L}$ & $1$ & $i$ & $0$ & $3$\\
           $N_{R}$ & $1$ & $i$ & $0$   & $3$\\
           \hline
            $L$& $1$ & $i$ &$0$  & $3$   \\
            $e_R$& $1$ & $i$&  $0$ &  $3$   \\
            \hline
            $\sigma$ & $1$ &$-1$ & 0 &1\\
			\hline
\end{tabular}
\caption{Field content of the dark sector and SM leptons. We impose the gauged $\text{SU}(3)_\text{D}$ and discrete $\mathcal{Z}_4$, whereas $\text{U}(1)_{\text{D}}$ is a residual symmetry present only after  the scalar $\sigma$ condenses. All dark fields are SM singlets.}
\label{tab:fields}
\end{table}

In order to draw as much as possible from our knowledge of QCD,  we choose the confining gauge group $\text{SU}(3)_\text{D}$ under which only the single  vector-like pair of dark quarks $(q_L, q_R)$ transform in the fundamental representation. We further add three generations of vector-like gauge singlet neutrinos $(N_L,N_R)$ and the real\footnote{This simplification is just in order avoid the presence of a Majoron \cite{Gelmini:1980re,Chikashige:1980ui} and not crucial to our mechanism.} scalar $\sigma$.
A $\mathcal{Z}_4$ symmetry forbids bare Majorana masses for $N_L\;(N_R)$ as well as a bare mass term for the dark quarks, (unlike the model in Ref.~\cite{Mitridate:2017oky}). The real singlet scalar $\sigma$ transforms as a $-1$ under  $\mathcal{Z}_4$ and has a positive mass squared $m_\sigma^2>0$. 
The relevant terms read
\begin{align}
   - \mathcal{L}_{LN} &= y_{e}  \overline{L}H e_R + y_{\nu}  \overline{L}\tilde{H} N_R + M_D \overline{N_L}N_R + \text{h.c.}\,,\\
   - \mathcal{L}_\text{D} &=	y_Q \sigma \overline{q_L} q_R + y_{N_L}  \sigma \overline{N_L^c} N_L\\
    &\quad + y_{N_R}  \sigma \overline{N_R^c} N_R + \text{h.c.}\,,\nonumber\\  V_{\sigma} &= \left(m_\sigma^2+\lambda_\sigma \sigma^2 + \lambda_{H\sigma} |H|^2 \right) \sigma^2\,,
\end{align}
where we suppressed flavor indices and $H$ is the SM Higgs doublet ($\tilde{H}\equiv i \sigma_2 H^*$) with a vev $v_H$ for its neutral component and all relevant charges and representations are summarized in Table~\ref{tab:fields}.
Gauge confinement of $\text{SU}(3)_\text{D}$ generates a quark condensate 
\begin{align}
\braket{\overline{q_L} q_R} \simeq \Lambda_\text{D}^3\,,
\end{align}
where $\Lambda_\text{D}$ is the confinement scale of the dark sector.\footnote{RG running of the dark gauge coupling $\alpha_{\text{D}}=g_{\text{D}}^2/4\pi$ is given at one loop by
\begin{equation*}
    \frac{1}{\alpha_{\text{D}}(E_2)}=\frac{1}{\alpha_{\text{D}}(E_1)}+\frac{\beta_0}{2\pi}\log\left(\frac{E_2}{E_1}\right), \qquad \beta_0=\frac{11}{3}N_c-\frac{2}{3}=\frac{31}{3}. 
\end{equation*}
The confinement scale is defined as the energy at which the dark gauge coupling turns non-perturbative, $\alpha_\text{D}(\Lambda_\text{D})\simeq 4\pi$.
}
This induces a vacuum expectation value for the new scalar $\sigma$
\begin{equation}\label{eq:vevsigma}
	\braket{\sigma}\simeq y_Q\frac{\Lambda_\text{D}^3}{m_\sigma^2}\,.
\end{equation}
Such an \enquote{induced} vev for $\sigma$ is reminiscent of the well known Type II seesaw \cite{Lazarides:1980nt,Schechter:1980gr,Mohapatra:1980yp,PhysRevD.22.2860,Wetterich:1981bx,Ma:2000cc,Davidson:2009ha}. In our study we typically obtain $\braket{\sigma} \ll m_H \ll m_\sigma$, where $m_H$ is the SM Higgs mass, so we can neglect mixing in the Higgs sector even for non-negligible $\lambda_{H\sigma}$. 
A non-vanishing $\braket{\sigma}$ implies   Majorana masses for $N_{L,R}$ from the Feynman diagram shown in Fig.~\ref{fig:numass}
\begin{equation}
	\mu_{L,R} \simeq  \SI{10} {\kilo\electronvolt}\cdot  y_Q y_{N_{L,R}} \left(\frac{\Lambda_\text{D}}{\SI{10}{\tera\electronvolt}}\right)^3 \left(\frac{\SI{3.1e8}{\giga\electronvolt}}{m_\sigma}\right)^2.
\end{equation}
Note that the inverse seesaw requires only one of the two mass terms $\mu_{L,R}$ to be present
to induce a small mass for $m_\nu$; if both terms are non-zero one obtains that at leading order in the seesaw expansion $\mu_{L,R} \ll y_{\nu} v_H \ll M_D$ \cite{CentellesChulia:2020dfh} 
\begin{align}
    m_\nu \simeq \frac{ \mu_L  y_{\nu}^2  v_H^2}{M_D^2 -\mu_L \mu_R}\,.
\end{align}
For the remainder of this work we define $\mu \equiv  \mu_L $ and ignore the subleading correction in the denominator, hence we use $m_\nu \simeq \mu\; (y_\nu v_H/M_D)^2$. The small Majorana masses $\mu_{L,R}$ are responsible for splitting the masses of the two chiralities of $N$ so that they become \textit{Pseudo-Dirac} fermions.

The spontaneous breaking of the  $\mathcal{Z}_4$ symmetry leads to the formation of domain walls. 
To avoid that we break this symmetry explicitly in the scalar potential by the bias term $\kappa \sigma|H|^2$ \cite{Sikivie:1982qv}. We assume a negligible trilinear term $\mu_{3} \sigma^3$ and   the linear piece $\mu_{1}^3 \sigma$ can always be set to zero by a field redefinition \cite{Espinosa:2011ax}. The dimensionful coupling $\kappa$ does not induce a sizeable shift to $\braket{\sigma}$ in Eq.~\eqref{eq:vevsigma} as long as 
\begin{align}
\kappa \ll \SI{3e7}{\giga\electronvolt}\cdot y_Q \left(\frac{\Lambda_\text{D}}{\SI{10}{\tera\electronvolt}}\right)^3,
\end{align}
and the domain walls decay before Big Bang Nucleosynthesis (BBN) for 
\begin{align}
  \kappa \gtrsim  \SI{e-13}{\giga\electronvolt}\cdot y_Q   \left(\frac{\Lambda_\text{D}}{\SI{10}{\tera\electronvolt}}\right)^3 \left(\frac{10^8\;\text{GeV}}{m_\sigma}\right).  
\end{align}
We checked that decays before BBN  occur long before the domain walls dominate the energy budget of the universe. Another way to remove the domain walls could be to invoke the fact that the $\mathcal{Z}_4$ symmetry is anomalous with respect to $\text{SU}(3)_\text{D}$ \cite{Preskill:1991kd}, but that requires a dedicated study.

\section{Particle spectrum}\label{sec:spec}
At temperatures below $\Lambda_\text{D}$ the theory confines and the dark quarks reorganize into dark hadrons. Since unlike in real QCD we work with only one generation of dark quarks there are no spontaneously broken chiral symmetries and hence no Nambu-Goldstone modes similar to the pions \cite{Francis:2018xjd}. The only global symmetries would be
\begin{align}
    \text{U}(1)_\text{D}\otimes \text{U}(1)_\text{A},
\end{align}
where the vectorial $\text{U}(1)_\text{D}$ is the dark equivalent of baryon number and the axial symmetry $\text{U}(1)_\text{A}$ is both explicitly broken by the coupling to $\sigma$ and anomalous to begin with.
We denote the lightest meson state $\ket{\overline{q}q}$ as $\mathcal{M}$, which is parity odd \cite{DellaMorte:2023ylq}.
Since the coupling to $\sigma$ explicitly breaks a chiral symmetry, one expects a dark quark mass of $m_Q\simeq y_Q \braket{\sigma}$ \cite{Hur:2011sv}. Due to the absence of spontaneously broken chiral symmetries and thus dark pions, the equivalent of the  Gell-Mann-Oakes-Renner relation \cite{Gell-Mann:1968hlm} $m_\pi \sim \sqrt{m_Q \Lambda_\text{D}}$ is not valid for the meson $\mathcal{M}$ and to reduce the number of free parameters we fix
\begin{align}
    m_\mathcal{M} \simeq \Lambda_\text{D}\,.
\end{align}
If we added more than one flavor of dark quarks we would find that the resulting lighter dark pions would typically be so long-lived that the injection of electromagnetic radiation from the electroweak showers of their leptonic decay products would alter BBN significantly \cite{Hambye:2021moy}  or they could be stable enough to overclose the universe.\footnote{From the Gell-Mann-Oakes-Renner relation the dark pions would get a tiny mass $m_\pi\simeq y_Q \Lambda_\text{D}^2/m_\sigma$ typically below $M_D$. Thus they would have to decay into active neutrinos with a highly suppressed rate $\sim m_\nu^2\Lambda_\text{D}^4/m_\sigma^5$.
A possible way to generalize our model to $N_f\geq2$ generations of dark quarks is to introduce a bare quark mass term, which sofly breaks the $\mathcal{Z}_4$ symmetry. This generate a larger mass for the dark pions above $M_D$ allowing them to decay into $N$ before BBN.}

The lightest dark baryon $\mathcal{B}$ consisting of $\ket{qqq}$ has spin $3/2$ \cite{Antipin:2015xia,Garani:2021zrr}, similar to the $\Delta$-resonance of the strong interaction, and its mass is expected from large-$N_c$ (number of dark colors) arguments to scale as $N_c \Lambda_\text{D}$ \cite{Antipin:2015xia}. In analogy to QCD, Ref.~\cite{Garani:2021zrr} finds a mass scale of about  $10 \Lambda_\text{D}$ and we interpolate between these two estimates by setting  
\begin{align}
    m_\mathcal{B} \simeq 5 \Lambda_\text{D}\,.
\end{align}
The stability of $\mathcal{B}$ at the renormalizable level is ensured by the conservation of dark baryon number $\text{U}(1)_\text{D}$ and hence it may be a good DM  candidate. 

For $\text{SU}(3)_\text{D}$ there can be glueball states that are odd or even under the dark sector charge conjugation. The even glueballs can   decay into two mesons as long as their mass is  above $2 m_\mathcal{M}$ in analogy to what is expected for QCD glueballs \cite{Ochs:2013gi,Cohen:2014vta}.
Odd glueballs might be stable and could be produced by similar dynamics to the dark baryons (see the next Section~\ref{sec:cosmo}), forming a component of  DM \cite{Buttazzo:2019iwr,Landini:2020daq,Gross:2020zam}. In the absence of more detailed knowledge of the mass spectrum for the case at hand, which would certainly require a lattice simulation, we focus on the dark baryon as DM and assume that the odd glueballs decay away. Even if these glueballs were stable, they would correct our results by only $\mathcal{O}(1)$ factors.\footnote{Glueballs have a typical mass $m_{\text{DG}}\simeq \Lambda_\text{D}$ and annihilate with cross section $\sigma_\text{DG}\simeq 1/\Lambda_\text{D}^2$, leading to qualitatively the same dynamics as for the dark baryons.}
If we chose the adjoint of $\text{SU}(3)_\text{D}$ for the representation of the dark quarks, there could be exotic hybrid bound states of dark quarks and dark gluons playing the role of DM \cite{Contino:2018crt}. 

\section{Cosmological History}\label{sec:cosmo}

We set both the reheating temperature $T_\text{RH}$ and the maximum temperature during reheating $T_\text{max}$, which can be much larger than $T_\text{RH}$ \cite{Giudice:2000ex,Kolb:2003ke} for non-instantaneous reheating, to be smaller than $m_\sigma$ so that we can safely integrate $\sigma$ out and treat it purely as a mediator.
Reheating sets the stage for a thermalized SM plasma and we assume that $T_\text{RH}\gg M_D,\Lambda_\text{D}$. As long as the  $N$ are relativistic, they are produced via their sizeable Yukawa coupling to the SM leptons at temperatures below
\begin{align}\label{eq:Nin}
     T_N^\text{in} \simeq
    \SI{7e7}{\giga\electronvolt}\cdot \left(\frac{y_\nu}{10^{-4}}\right)^2 \sqrt{\frac{106.75}{g_{*\rho}(T_N^\text{in})}}\,,
\end{align}
where $g_{*\rho}$ is the number of relativistic degrees of freedom in  the energy density and they consequently populate the dark sector via their fast $\sigma$-mediated annihilations into dark quarks $NN\leftrightarrow\overline{q}q$ at $T>\Lambda_\text{D},M_D$.
Dark gluons are produced from the thermalized quarks via $\overline{q}q \leftrightarrow gg$
with a rate of about $g_\text{D}^4 T$. Requiring that this comes into thermal equilibrium above the dark confinement scale amounts to the requirement
\begin{align}
    \Lambda_\text{D} \lesssim 10^{10}\;\text{GeV} \cdot \left(\frac{g_\text{D}}{0.01}\right)^4 \sqrt{\frac{106.75}{g_{*\rho}(T_{gg}^\text{in})}}\,,
\end{align}
which, as we will see in Eq.~\eqref{eq:DM}, is always satisfied, especially for non-perturbative $g_\text{D}$.

The abundance of dark baryons is determined via thermal freeze-out of the annihilations $\overline{\mathcal{B}}\mathcal{B}\leftrightarrow \overline{\mathcal{M}} \mathcal{M}$, which occurs in the $s$-wave \footnote{Following the argument in Refs.~\cite{Contino:2018crt,Gross:2018zha}, the maximum angular momentum of the annihilation process can be estimated as $l= \mu |\vec{v}|b$, where $b\simeq 1/\Lambda_\text{D}$ is the impact parameter, $|\vec{v}|\simeq \sqrt{T/m_\mathcal{B}}$ the relative velocity and $\mu=m_\mathcal{B}/2$ is the reduced mass. As $l\simeq \sqrt{T m_\mathcal{B}}/(2 \Lambda_\text{D})$ we find $l<1$ both at the time of DM freeze-out ($l\simeq 1/2$ see Eq.~\eqref{eq:BFO}) and today, so the relevant processes always take place in the fully quantum regime. In this regime, the $s$-wave contribution is expected  to dominate the total annihilation cross section $\braket{\sigma_\text{D} |\vec{v}|}$ \cite{Contino:2018crt}. Notice that in \cite{Contino:2018crt,Gross:2018zha} some of the processes take place in the semiclassical regime because they consider hadrons made of heavy quarks. }  with a geometric cross section  \cite{Buttazzo:2019iwr}  of \cite{Steigman:2012nb}
\begin{align}\label{eq:cross}
    \braket{\sigma_\text{D} |\vec{v}|} \simeq  \frac{\pi}{\Lambda_D^2} \simeq  2.2 \times 10^{-26}\, \frac{\text{cm}^3}{\text{s}} \cdot \left(\frac{\SI{41}{\tera\electronvolt}}{\Lambda_\text{D}}\right)^2.
\end{align}
Here we neglect any potential enhancement of the cross section due to intermediate resonances, as occurs e.g. for the proton due to the deuteron resonance \cite{Cline:2013zca}.
To logarithmic accuracy we find that these annihilations decouple at 
\begin{align}\label{eq:BFO}
    T_{\mathcal{B}\mathcal{M}}^\text{out} \simeq \frac{m_\mathcal{B}}{25}
\end{align}
and that the relic abundance in general is reproduced for values of \cite{Antipin:2015xia,Garani:2021zrr}
\begin{align}\label{eq:DM}
    \Lambda_\text{D} \simeq (1-100)\;\text{TeV}.
\end{align}
We stress that, due to $\mathcal{O}(1)$ uncertainties in the hadron spectrum, as well as in the thermally-averaged cross section of Eq.~\eqref{eq:cross}, we are unable to determine the precise value of $\Lambda_{\text{D}}$ that reproduces the relic abundance and can only estimate a reasonable range in Eq.~\eqref{eq:DM}.
Even though the dark baryon is self-interacting with $\sigma_\mathcal{B}/m_\mathcal{B} \sim 1/\Lambda_\text{D}^3$, the above range for $\Lambda_\text{D}$ precludes a strong enough elastic cross section that would be necessary to solve the \enquote{cusp-core}- \cite{Borriello:2000rv,2009MNRAS.397.1169D,2010AdAst2010E...5D,deBlok:2002vgq} or  \enquote{too-big-to-fail}-problems \cite{2011MNRAS.415L..40B,2012MNRAS.422.1203B} (for an overview see Ref.~\cite{Cline:2013zca} and references within), which would require sub-GeV values of $\Lambda_\text{D}$. For the same reason our scenario is not constrained by bounds from the halo elipticity \cite{Miralda-Escude:2000tvu} or the Bullet Cluster \cite{Randall:2008ppe}.

\begin{figure*}
    \includegraphics[width=0.49\textwidth]{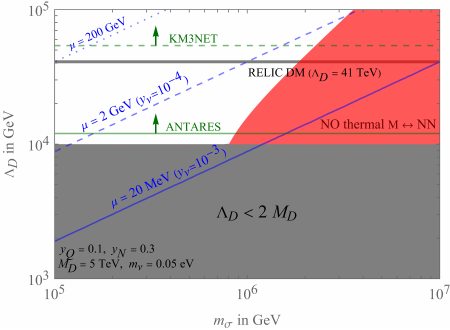}
    \includegraphics[width=0.49\textwidth]{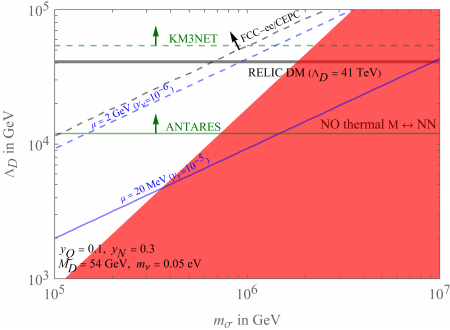}
     \caption{Allowed parameter space for the successful production of the dark matter together with iso-contours for $\mu$ and the corresponding $y_\nu$ required to explain $m_\nu = \SI{0.05}{\electronvolt}$ in the inverse seesaw mechanism for two sets of benchmark parameters with TeV-scale \textit{(left)} and GeV-scale \textit{(right)} $N$. The green lines indicate the current and projected lower limits  on $\Lambda_\text{D}$ from indirect detection via dark matter annihilations into dark mesons, followed by a chain of two-body decays producing neutrinos. For $N$ around the GeV-scale the parameter space might also be probed by displaced vertex searches at future colliders.}
    \label{fig:plot1}
\end{figure*}

A crucial ingredient of the freeze-out estimate is that the dark sector maintains the same temperature as the SM bath.
Indeed, below the confinement scale all the dark hadrons are non-relativistic. Therefore, in absence of thermal contact with the SM, the temperature of the dark thermal bath would red-shift only logarithmically with the scale factor, analogously to \enquote{cannibal} DM models \cite{Pappadopulo:2016pkp,Farina:2016llk}, leading to an over-abundant population of dark baryons.  Thermal equilibrium (between both sectors\footnote{An alternative possibility is that the dark sector particles thermalize among themselves, but not with the SM, forming a secluded dark bath which evolves with its own temperature $T_d\equiv\xi T$ \cite{Morrison:2020yeg}. In such a case the DM evolution depends on the additional free parameter $\xi$, which encodes some unknown initial condition, so that $\xi\simeq 10^{-3}$ is needed to reproduce the correct relic abundance for TeV-scale DM. We do not consider this possibility here.} ) below the confinement scale is maintained via decays and inverse decays $\mathcal{M}\leftrightarrow \overline{N} N$, because the $N$ are tightly coupled with the SM at temperatures above their mass $M_D$, see Eq.~\eqref{eq:Nin}. One finds a decay rate of 
\begin{align}
    \Gamma( \mathcal{M}\rightarrow \overline{N} N)\simeq \frac{y_Q^2 y_N^2}{32 \pi} \frac{m_\mathcal{M} f_\mathcal{M}^2}{m_\sigma^4}\sqrt{1-\frac{4M_D^2}{m_{\mathcal{M}}^2}}\,,
\end{align}
in terms of the matrix-element $f_\mathcal{M}$, that we parameterize as 
\begin{align}
    f_\mathcal{M}\equiv \bra{0}\overline{q}\gamma_5 q\ket{\mathcal{M}}\simeq \Lambda_\text{D}^2\,,    
\end{align}
and $y_N\equiv y_{N_L}+y_{N_R}$. 
By employing Maxwell-Boltzmann statistics we find that the thermally averaged decay rate for $\mathcal{M}\rightarrow \overline{N} N$ reads \cite{Davidson:2008bu}
\begin{align}
    \braket{\Gamma_\text{D}} = \frac{\text{K}_1\left(\frac{m_\mathcal{M}}{T}\right)}{\text{K}_2\left(\frac{m_\mathcal{M}}{T}\right)} \Gamma( \mathcal{M}\rightarrow \overline{N} N)
\end{align}
in terms of the modified Bessel functions of the second kind $\text{K}_{1,2}\left(m_\mathcal{M}/T\right)$.
From the principle of detailed balance we obtain for the thermal average of the inverse decay rate $\overline{N} N\rightarrow \mathcal{M}$,
\begin{align}
    \braket{\Gamma_\text{ID}} = \frac{n_\mathcal{M}^\text{eq.}}{n_{N}^\text{eq.}} \braket{\Gamma_\text{D}}.
\end{align}
Here we introduced the equilibrium number density of particles species $i$ with mass $m_i$ and $g_i$ internal degrees of freedom via 
\begin{align}
    n_i^\text{eq.} = \frac{g_i T^3}{2\pi^2} \left(\frac{m_i}{T}\right)^2 \text{K}_2\left(\frac{m_i}{T}\right),
\end{align}
 with $g_\mathcal{M}=1,\; g_N=2$. We require that that the inverse decay remains in equilibrium until at least the temperature of the $\mathcal{B}$ freeze-out defined in Eq.~\eqref{eq:BFO}. On top of that we impose $m_\mathcal{M}=\Lambda_\text{D}>2 M_D$ so that the decay channel is kinematically open.  Further we check that the dark matter relic density is not diluted by entropy release: we find that decay $\mathcal{M}\rightarrow \overline{N} N$ is always fast, as expected from the previous arguments, and that the decay width of $N$ given by  $\Gamma(N\rightarrow L H)\simeq y_\nu^2 M_D/(8\pi)$ equals the Hubble rate before the temperature $T_\text{dom}\simeq 7 M_D / (4 g_{*\rho}(T_\text{dom}))$ \cite{Giudice:1999fb} (when $N$ would start to dominate over the energy density of radiation) as long as 
\begin{align}\label{eq:ooedec}
    y_\nu \gtrsim 7.8\times 10^{-9}\cdot \sqrt{\frac{M_D}{10\;\text{TeV}}} \left(\frac{g_{*\rho}(T_\text{dec})}{106.75}\right)^\frac{1}{4} \left(\frac{106.75}{g_{*\rho}(T_\text{dom})}\right).
\end{align}
In Fig.~\ref{fig:plot1} we depict the parameter space in the $\Lambda_\text{D}$ versus $m_\sigma$ plane subject to the previously discussed constraints. 
For the showcased benchmark point we obtain the observed relic density of $\Omega_\mathcal{B}h^2=0.120\pm 0.001$ \cite{Planck:2018vyg} by using Eq.~\eqref{eq:cross} and values of  $\mu > \mathcal{O}(\SI{100}{\mega\electronvolt})$ corresponding to $y_\nu < \mathcal{O}(10^{-3})$, bounded from below by Eq.~\eqref{eq:ooedec}.
These values for the lepton number breaking parameter are larger than the conventionally assumed keV-scale (see e.g. Eq.~\eqref{eq:ISS}), but bear the additional advantage that $M_D = \mathcal{O}(\text{TeV})$, which is in range of future collider experiments. One can understand the largeness of $\mu$ from the plot in Fig.~\ref{fig:plot1} by noting that for smaller values of $\mu$ (equivalent to larger $y_\nu$ for fixed $m_\nu$) the relic abundance would only be reached in the red colored region where $\mathcal{M}$ is not thermalized long enough.  
For values of $\mu$ above the scale of $M_D$ the   inverse seesaw expansion breaks down and we would be in the usual Type I Seesaw regime.
A smaller value for the product $y_Q y_N$  moves the $\mu$-isocontours upwards along the $\Lambda_\text{D}$ axis and increases the size of the region excluded by the meson thermalization.  
Finally, let us point out that our scheme involving TeV-scale $N$ might  reproduce the baryon asymmetry of the universe \cite{Agashe:2018cuf} via resonantly enhanced \cite{Pilaftsis:2003gt,Pilaftsis:2004xx,Pilaftsis:2005rv} out-of-equilibrium decays. 

\section{Signals \& constraints
}\label{sec:signals}

\begin{figure*}
    \includegraphics[width=0.6\textwidth]{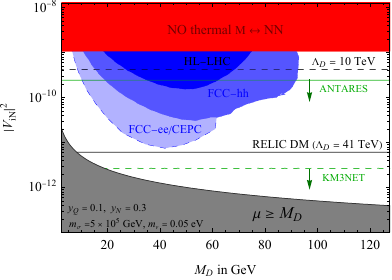}
     \caption{Sensitivity projections of displaced vertex searches for GeV-scale $N$ together with limits and projections from neutrino telescopes 
     in the parameter space that is compatible with our cosmological considerations.    Our cosmological analysis is independent of flavor and the collider limits are for the electron and muon channels. In the gray region the inverse seesaw scheme breaks down as the Majorana masses $\mu$ of $N$ become larger than their Dirac masses $M_D$.}
    \label{fig:plot2}
\end{figure*}

For our parameter space with $\mu > \mathcal{O}(\SI{10}{\mega\electronvolt})$ the mixing between active neutrinos of flavour $i=e,\mu,\tau$ and $N$ scales as $|V_{iN}|^2 \simeq m_\nu /\mu \lesssim \mathcal{O}\left(10^{-[9,8]}\right)$. A recent review of all pertinent laboratory constraints was compiled in Ref.~\cite{Bolton:2019pcu}: For $M_D=\mathcal{O}(\text{TeV})$ the  constraint from electroweak precision observables due to the modification of charged- and neutral-current reactions induced by the non-unitarity in the active neutrino sector   reads $|V_{iN}|^2 < 10^{-3}$ \cite{Abada:2007ux,Fernandez-Martinez:2016lgt,Blennow:2016jkn}. Next generation electron colliders such as FCC-ee or CEPC could improve the non-unitarity  bound for the mixing with electron-neutrinos down to $|V_{eN}|^2 < 10^{-[5,4]}$ \cite{Antusch:2016ejd}. Displaced vertex searches for $N$ masses below about $\SI{100}{\giga\electronvolt}$ at the high luminosity upgrade of the LHC might probe values down to $|V_{iN}|^2 \simeq 5\times 10^{-10}$, while the proposed hadronic collider FCC-hh might reach $|V_{iN}|^2 \simeq 5\times 10^{-11}$  \cite{Antusch:2016ejd}, whereas FCC-ee or CEPC could potentially test mixings as small as $|V_{iN}|^2 \simeq 10^{-11}$ \cite{Antusch:2016vyf,Blondel:2022qqo,Abdullahi:2022jlv}.
These searches for long lived  $N$ far below the TeV-scale could test our cosmologically preferred parameter space (see the right plot in Fig.~\ref{fig:plot1}), as can be observed in Fig.~\ref{fig:plot2}.

When it comes to charged lepton flavor violation the strongest constraints come from the non-observation of the decay $\mu\rightarrow e \gamma$ at MEG \cite{MEG:2016leq} and MEG II \cite{MEGII:2018kmf} setting a combined limit of $3.1\times10^{-13}$ \cite{MEGII:2023ltw},
and non-observation of muon-to-electron-conversion on titanium by the SINDRUM II collaboration with an upper limit on the branching ratio for ground state transitions of $R_{\mu \rightarrow e}^\text{Ti}< 4.3\times 10^{-12}$ \cite{SINDRUMII:1993gxf}, that impose $\left|V_{e N} V_{\mu N}^*\right|<10^{-5}$ in the range $\SI{100}{\giga\electronvolt}<M_D<\SI{10}{\tera\electronvolt}$ \cite{Bolton:2019pcu}. At two loops in the inverse seesaw \cite{deGouvea:2005jj,Abada:2015trh} there are contributions to the electric dipole moment of the electron, which is measured by the ACME II experiment to be  
$|d_e| < 1.1\times 10^{-29} \;e\;\text{cm}$ \cite{ACME:2018yjb} and in the future this result is expected to improve by about an order of magnitude. The authors of Ref.~\cite{Abada:2016awd} found for $M_D=\mathcal{O}(\text{TeV})$ the largest possible value of $|d_e|=10^{-[32,31]}\;e\;\text{cm}$, which could be testable in the future. 
For the muon electric dipole moment the current direct limit of $|d_\mu| < 1.9\times 10^{-19} \;e\;\text{cm}$ \cite{Muong-2:2008ebm} was obained at Brookhaven National Laboratory (BNL)  and indirect limits from heavy atoms and molecules via muon loops reach down as far as $2\times 10^{-20} \;e\;\text{cm}$ \cite{Ema:2021jds}, whereas 
future experiments at PSI and  J-PARC are expected to improve these bounds to $6\times 10^{-23} \;e\;\text{cm}$ \cite{Adelmann:2021udj,Sakurai:2022tbk,muonEDMinitiative:2022fmk} and $10^{-24} \;e\;\text{cm}$ \cite{Feng:2001mq} respectively. By rescaling the result from Ref.~\cite{Abada:2016awd} for the electron electric dipole moment in the inverse seesaw by a factor of $m_\mu / m_e$ we find   
$|d_\mu|\simeq 2\times 10^{-[30,29]}\;e\;\text{cm}$,
which is out of reach of   future experiments.
In Ref.~\cite{Pinheiro:2021mps} it was concluded that the inverse seesaw cannot account for the  discrepancy in the anomalous magnetic dipole moment of the muon observed by BNL \cite{Muong-2:2021ojo}. 

For $N$ above the GeV-scale one expects that the dominant contribution to neutrinoless-double-beta decay comes from the exchange of the light neutrinos. An estimate of the resulting rate depends on assumptions about flavor, and for the normal hierarchy of active neutrino masses there remains the possibility of an accidental cancellation drastically reducing the rate below future sensitivities depending on the interplay of so far unknown lightest neutrino mass, the observed mixing angles and the possible Majorana phases in the PMNS matrix \cite{Denton:2023hkx}.

There will not be a signal in the gravitational wave spectrum from the minimal dark confinement transition considered here, since for three dark colors and one dark flavor one expects a smooth crossover (see Ref.~\cite{Bai:2018dxf} and references within).
Our setup is not constrained by direct detection of DM due to the smallness of the mixing between $\sigma$ and the SM Higgs boson (see the discussion below Eq.~\eqref{eq:vevsigma}).

Indirect detection on the other hand, offers the intriguing prospect of signals at future neutrino telescopes: 
Annihilations of $\mathcal{B}$ into mesons followed by the immediate decay of $\mathcal{M}$ into two TeV-scale $N$, which in turn is followed by the decays $N\rightarrow \sum_i H_0 \nu_i, \;\sum_i Z_\mu \nu_i, \;  \sum_i W_\mu^\mp l_i^\pm$ with $\text{BR}(N\rightarrow \sum_i H_0 \nu_i)=\text{BR}(N\rightarrow \sum_i Z_\mu \nu_i)=\text{BR}(N\rightarrow  \sum_i W_\mu^\mp l_i^\pm)/2$ \cite{Das:2018usr},    would produce monoenergetic primary neutrinos with an energy of $m_\mathcal{B}/4$ (assuming $m_\mathcal{B}\gg m_\mathcal{M}$ for simplicity) plus less energetic secondary neutrinos and a background of SM particles   from the decays and interactions of the $H_0, Z_0, W^\mp, l^\pm$ also present.\footnote{For the case of light $N$ with masses below the electroweak scale the available decay modes of $N$ depend on its mass, which will affect the number of produced neutrinos and their energy spectra. We do not expect the order of magnitude estimates obtained for TeV-scale $N$ to drastically change in this limit. } For reactions producing initially monoenergetic neutrinos with energies above $\mathcal{O}(\SI{100}{\giga\electronvolt})$, there would never be an exactly monochromatic neutrino line as the produced neutrinos would be so energetic that they would produce electroweak bremsstrahlung resulting in cascades similar to QCD jets at colliders \cite{Berezinsky:2002hq,Kachelriess:2009zy,Ciafaloni:2010ti}. Reference \cite{ElAisati:2015ugc} concluded that the widening of the line is not larger than the energy resolution of high energy neutrino telescopes \cite{IceCube:2013dkx} for dark matter masses in the range $10^{[3,8]}\;\text{GeV}$, which is why in the following we neglect this effect. 

Current and projected bounds on neutrinos produced in dark matter annihilations were compiled in Ref.~\cite{Arguelles:2019ouk} and for the $m_\mathcal{B}=\mathcal{O}(10^{[4,5]} \,\text{TeV})$ mass range the strongest bound (for $s$-wave annihilations) for neutrino telescopes of $\braket{\sigma |\vec{v}|}< \left(5\times 10^{-24}-10^{-23}\right)\; \text{cm}^3/\text{s}$ was obtained by the ANTARES collaboration \cite{ANTARES:2015vis}. This limit is expected to improve to $10^{-24}\; \text{cm}^3/\text{s}$ \cite{IceCube:2019pna} for the expansion of the current IceCube observatory \cite{IceCube:2014gqr}, $\left(10^{-25} -5\times 10^{-25}\right)\; \text{cm}^3/\text{s}$  for the proposed Pacific Ocean Neutrino experiment \cite{P-ONE:2020ljt} and $5\times10^{-[26,25]}\; \text{cm}^3/\text{s}$ for the KM3NeT \cite{KM3Net:2016zxf,KM3NeT:2018wnd} water Cherenkov detector currently being constructed with a $\text{km}^3$ volume in the Meditarrenean sea. From  the aforementioned electroweak showers and the charged SM particles   that can also be emitted  in the $N$ decays we also expect that high energy photons should be produced, which is constrained by gamma ray data from Fermi-LAT \cite{Fermi-LAT:2015att} and HESS \cite{HESS:2016mib} leading to $\braket{\sigma |\vec{v}|}<  10^{-23}\; \text{cm}^3/\text{s}$ \cite{Queiroz:2016zwd,Batell:2017rol}
and will be probed further by the upcoming Cherenkov Telescope Array \cite{Morselli:2017ree} with a projected limit of  $\left( 10^{-24}-5\times10^{-24}\right)\; \text{cm}^3/\text{s}$ \cite{Queiroz:2016zwd,Batell:2017rol}. Note that a detailed limit for our case will depend on the energy fraction deposited in the photons, which requires a dedicated simulation of the decay chain and showering \cite{Batell:2017rol}.

All of the above limits assume spin 1/2 Majorana dark matter (2 degrees of freedom), but since we we have a spin 3/2 Dirac fermion ($2\cdot (2\cdot 3/2 + 1)=8$ degrees of freedom) we need to rescale the limits by a factor of $1/4$. Also limits from annihilations assume a neutrino energy equal to the dark matter mass, but for our case the right energy range is roughly $m_\mathcal{B}/4$. 
We estimate the thermally averaged dark matter annihilation cross section by using \eqref{eq:cross} and with the rescaled limit from ANTARES \cite{ANTARES:2015vis} we find 
\begin{align}
    \Lambda_\text{D} > \SI{12}{\tera\electronvolt} \cdot \sqrt{ \frac{0.25\times 10^{-24}\; \text{cm}^3/\text{s}}{\braket{\sigma |\vec{v}|}}}\,,
\end{align}
which will improve to $\SI{54}{\tera\electronvolt}$ once  KM3NeT \cite{KM3Net:2016zxf,KM3NeT:2018wnd} is operational so that our benchmark of $\Lambda_\text{D}= \SI{41}{\tera\electronvolt}$ should be tested by next generation experiments; we demonstrate the impact on our parameter space in Fig.~\ref{fig:plot1}. Of course we should stress that this is just an order of magnitude estimate due the uncertainties related to the non-perturbative dynamics of the dark hadrons,
and setting $\text{BR}(\mathcal{M}\rightarrow \overline{N} N) \simeq 1$.

If DM is stabilized by a global symmetry such as our  accidental $\text{U(1)}_\text{D}$, there might be higher dimensional operators from the putative field theoretic UV-completion that break this symmetry, or Planck-suppressed operators due to non-perturbative quantum gravitational effects, which are expected to violate all global symmetries \cite{COLEMAN1988643,GIDDINGS1988854,GILBERT1989159,Kallosh:1995hi}, and thus induce DM decay \cite{Mambrini:2015sia}. We summarized the relevant effective operators for our case in appendix \ref{sec:high-dim} and find that the lowest dimensional ones occur at dimension $d=8$ in Eqs.~\eqref{eq:eight}-\eqref{eq:eight2}. On dimensional grounds we parameterize 
\begin{align}
   f_\mathcal{B} \equiv \bra{0} qqq\ket{\mathcal{B}} \simeq \Lambda_\text{D}^3\,.
\end{align}
The operator in equation \eqref{eq:eight} induces the decays $\mathcal{B}_\mu \rightarrow \sum_i H_0 \nu_i$,
$\mathcal{B}_\mu \rightarrow \sum_i Z_\mu \nu_i$ and $\mathcal{B}_\mu \rightarrow \sum_i W_\mu^\mp l_i^\pm$. Owing to the fact that $m_\mathcal{B} \gg m_H,\;m_Z,\; m_W$ one finds that \cite{Garcia:2020hyo} $\Gamma\left(\mathcal{B}_\mu \rightarrow \sum_i H_0 \nu_i\right)= \Gamma\left(\mathcal{B}_\mu \rightarrow \sum_i Z_\mu \nu_i\right) = \Gamma\left(\mathcal{B}_\mu \rightarrow \sum_i W_\mu^\mp l_i^\pm\right)/2$ and their sum reads \cite{Garcia:2020hyo} 
\begin{align}\label{eq:rate1}
    \Gamma_2^{(1)}  = \frac{\left|c_8^{(1)}\right|^2 \Lambda_\text{D}^6}{\Lambda_\text{UV}^8} \frac{m_\mathcal{B}^3}{256 \pi}\,.
\end{align}
There also exist the three-body modes 
$\mathcal{B}_\mu \rightarrow \sum_i H_0 Z_\mu \nu_i $ and $\mathcal{B}_\mu \rightarrow \sum_i H_0 W_\mu l_i $, that are enhanced by a factor of $m_\mathcal{B}^2/v_H^2$ compared to the previous two-body decays and thus dominate over them. We estimate their sum to be 
\begin{align}
    \Gamma_3^{(1)} \simeq \frac{\left|c_8^{(1)}\right|^2 \Lambda_\text{D}^6}{\Lambda_\text{UV}^8} \frac{3 m_\mathcal{B}^5}{8192 \pi^2 v_H^2}\,.
\end{align}
In Ref.~\cite{ElAisati:2015ugc} it was found, that the energy spectrum for a three-body decay can be approximated by a power law with $\text{d}N/\text{d}E \sim (E/m_\mathcal{B})^{-[2,3]}$ for which they derive a limit of about $\tau_\mathcal{B}>10^{28}\;\text{s}$
in the window $m_\mathcal{B}=\mathcal{O}(10^4-10^5\;\text{GeV})$ using data from IceCube \cite{IceCube:2014rwe} and we obtain
\begin{align}\label{eq:bound1}
\frac{\Lambda_\text{UV}}{10^{12}\,{\rm GeV\,}}\gtrsim \left|c_8^{(1)}\right|^\frac{1}{4} \left(\frac{m_\mathcal{B}}{5 \Lambda_\text{D}}\right)^\frac{5}{8} \left(\frac{\Lambda_\text{D}}{\SI{40}{\tera\electronvolt}}\right)^\frac{11}{8} \left(\frac{10^{28}\;\text{s}}{\tau_\mathcal{B}}\right)^\frac{1}{8}.
\end{align}
The second operator in Eq.~\eqref{eq:eight2} leads to the decay modes $\mathcal{B}_\mu \rightarrow A_\mu N$ and $\mathcal{B}_\mu \rightarrow   Z_\mu N$ followed by the aforementioned two-body decay of $N$ to SM states. The total width  for our case of  $m_\mathcal{B}\gg M_D$ is found to be \cite{Garcia:2020hyo} 
\begin{align}\label{eq:rate2}
        \Gamma_2^{(2)} = \frac{\left|c_8^{(2)}\right|^2 \Lambda_\text{D}^6}{\Lambda_\text{UV}^8} \frac{m_\mathcal{B}^3}{4 \pi}\,.
\end{align}
Consequently we  expect  mono-chromatic neutrino lines  from the DM decays due the operator in Eq.~\eqref{eq:eight2},  while the previously mentioned, but negligible, widening of the lines due to electroweak cascades applies again. The authors of \cite{ElAisati:2015ugc}   derived a limit on the DM lifetime for two-body decays of about $\tau_\mathcal{B}>10^{28}\;\text{s}$ in the aforementioned window of DM masses. Applying this limit to the rate in \eqref{eq:rate2} leads to the constraint  
\begin{align}
\frac{\Lambda_\text{UV}}{6\times 10^{11}\,{\rm GeV\,}}>  \left|c_8^{(2)}\right|^\frac{1}{4} \left(\frac{m_\mathcal{B}}{5 \Lambda_\text{D}}\right)^\frac{3}{8} \left(\frac{\Lambda_\text{D}}{\SI{40}{\tera\electronvolt}}\right)^\frac{9}{8} \left(\frac{10^{28}\;\text{s}}{\tau_\mathcal{B}}\right)^\frac{1}{8},
\end{align}
which is slightly weaker than Eq.~\eqref{eq:bound1}, 
and the bound for the rate in Eq.~\eqref{eq:rate1} would be found by replacing $c_8^{(2)}$ with $c_8^{(1)}/8$.

\section{Conclusions}\label{sec:concl}
We have proposed a dark sector that dynamically generates the lepton-number breaking mass term of an electrically neutral fermion via the condensation of a single generation of dark quarks. This may be used to generate either the Majorana mass of the heavy right-handed neutrinos present in Type-I/III seesaw\footnote{In principle, any Majorana mass scale (say e.g. eV to $10^{14}$ GeV) could be realized. However, in order to generate very heavy Majorana masses, larger than $\mathcal{O}(100)$ TeV, and reproduce the relic abundance via freeze-out (which requires $\Lambda_{\text{D}} \lesssim \mathcal{O}(100)$ TeV) one needs $m_\sigma \ll \Lambda_{\text{D}}$, which is a different region of the parameter space to the one analyzed in this work.} or the small $\mu$ term present in the inverse seesaw mechanism. We focus on the latter possibility because it has a richer phenomenology. 

The resulting massive spin $3/2$ dark baryon which emerges from the confinement of the dark sector is analogous to the $\Delta$-baryon of QCD and  is stabilized by a dark baryon number symmetry, which is only violated by higher dimensional operators starting at dimension eight due to its larger spin. Thus it constitutes a good DM candidate and we obtain its yield  from thermal freeze-out in the dark sector. This model predicts a dark confinement scale of the order of $\Lambda_\text{D} = \mathcal{O}(1-100)\;\text{TeV}$ to reproduce the dark matter relic abundance  from which we find that $\mu > \mathcal{O}(\SI{10}{\mega\electronvolt})$ and a heavy neutrino mass of  $M_D=\mathcal{O}(\text{TeV})$. The neutrino portal from the inverse seesaw is crucial for keeping the dark hadrons in thermal contact with the SM. We have studied the phenomenological implications of the TeV-scale Pseudo-Dirac neutrinos and the potential signals from DM decays and annihilations, the latter of which may be probed by the upcoming KM3NeT experiment. Furthermore, a detailed analysis of the energy spectra of the decay products relevant for indirect detection would be interesting to pursue. 

In the case of GeV-scale sterile neutrinos, displaced vertex searches at proposed future colliders FCC-ee and FCC-hh may probe a significant part of the allowed parameter space.

\section{Acknowledgements}
We would like to thank Alessandro Strumia  for helpful discussions.
MB and JHG  are supported by \enquote{Consolidación Investigadora Grant CNS2022-135592}, funded also by \enquote{European Union NextGenerationEU/PRTR}. JHG is supported by the \enquote{Generalitat Valenciana} through the GenT Excellence Program (CIDEGENT/2020/020). G.L. is supported by the Generalitat Valenciana APOSTD/2023 Grant No. CIAPOS/2022/193. This work is partially supported by the Spanish \enquote{Agencia Estatal de Investigación} MICINN/AEI (10.13039/501100011033) grants PID2020-113334GB-I00 and PID2020-113644GB-I00.

\appendix

\section{Spin $3/2$ fermions and dark matter stability}\label{sec:high-dim}
Here we briefly review the properties of spin $3/2$  Rarita-Schwinger fields \cite{Rarita:1941mf} in order to construct the interactions of our DM candidate $\mathcal{B}\sim \ket{qqq}$ which is analogous to the $\Delta$-baryon of QCD.
We begin with a \enquote{spinor-vector} $\psi^\mu$, which is the direct product of the vector representation $(\frac{1}{2},\frac{1}{2})$ and the spinor representation of a Dirac fermion $(\frac{1}{2},0)\oplus (0,\frac{1}{2})$, which gives 
\begin{align}
    \left(1,\frac{1}{2}\right) \oplus \left(0,\frac{1}{2}\right)\oplus \left(\frac{1}{2},1\right) \oplus \left(\frac{1}{2},0\right),
\end{align}
and corresponds to 16 degrees of freedom (d.o.f).
By imposing the constraint on the free theory \cite{Weinberg:1995mt} that
\begin{align}
\gamma_\mu \psi^\mu=0 \label{eq:constr1},
\end{align}
one can eliminate the spin $1/2$ Dirac spinor  $(\frac{1}{2},0)\oplus (0,\frac{1}{2})$, which corresponds to four d.o.f. Further imposing a second constraint \cite{Weinberg:1995mt},
\begin{align}
       \partial_\mu \psi^\mu =0 \label{eq:constr2},
\end{align}
eliminates four more d.o.f from another spin $1/2$ Dirac spinor and the remaining eight physical d.o.f correspond to a spin $3/2$ Dirac fermion of mass $m$ that obeys the Dirac equation. It has a kinetic term given by
\begin{align}
    \overline{\psi}^\mu \Lambda_{\mu\nu}\psi^\nu
\end{align}
in terms of \cite{Haberzettl:1998rw}
\begin{align}
    \Lambda_{\mu \nu } &=  - \left(\slashed{p}-m\right) g_{\mu\nu} + A \left(\gamma_\mu p_\nu + p_\mu \gamma_\nu\right) \\
    &+ \frac{1}{2}\left(1+2A + 3 A^2 \right) \gamma_\mu \slashed{p} \gamma_\nu 
    + m \left(1 + 3 A + 3A^2\right) \gamma_\mu \gamma_\nu \nonumber.
\end{align}
Here $A$ is a free parameter with the requirement $A\neq -1/2$ to avoid a singular propagator  \cite{Haberzettl:1998rw} and the structure of the kinetic term can be obtained by requiring invariance under the following field redefinition \cite{Nath:1971wp,Benmerrouche:1989uc}
\begin{align}
    \psi^\mu &\rightarrow \left(g^{\mu \nu} + a \gamma^\mu \gamma^\nu\right) \psi_\nu,\\
    A&\rightarrow \frac{A-2a}{1+4 a},
\end{align}
in terms of another free parameter $a\neq -1/4$. 
One can think of $A$ as parameterizing  the admixture of the spin $1/2$ component $\gamma_\mu \psi^\mu$ in the off-shell $\psi^\mu$ field \cite{Nath:1971wp}.
Due tothe invariance of the Lagrangian under the above transformations the parameter $A$ will  drop out of all physical observables as shown by Ref.~\cite{Kamefuchi:1961sb}. By imposing invariance under the aforementioned field redefinitions  one can construct the interaction of $\psi^\mu$  with fermions and pseudo-scalars e.g.  the coupling of  $\Delta$ to pions $\pi$ and nucleons $n$ \cite{Nath:1971wp,Benmerrouche:1989uc}
\begin{align}
    \mathcal{L}_{\Delta\pi n} = c_{\Delta\pi n}\; \Delta^\mu \theta_{\mu\nu}\left(\partial^\nu \pi\right) n +\text{h.c.},
\end{align}
where we suppressed isospin and the electric charges and one defines \cite{Nath:1971wp,Benmerrouche:1989uc}
\begin{align}
    \theta_{\mu\nu}\equiv g_{\mu\nu} + \left(\frac{A}{2}(1+ 4 z) +z \right) \gamma_\mu \gamma_\nu,
\end{align}
where the free parameter $z$ is known as the \enquote{off-shell parameter}, that arises because the interaction involves the  spin $1/2$ components of the off-shell $\psi^\mu$.
In the context of chiral perturbation theory (see Ref.~\cite{Scherer:2002tk} for a review) one can absorb $z$ in the Wilson coefficients of certain contact terms via a field redefinition \cite{Tang:1996sq,Ellis:1996bd,Krebs:2009bf}, rendering it redundant.
In Supergravity theories (see Ref.~\cite{Nilles:1983ge} for a review) the elementary spin $3/2$ fermion known as the gravitino obtains its mass from a spin $1/2$ fermion known as the Goldstino via the super-Higgs mechanism \cite{Cremmer:1978iv},  so that it has a spin $1/2$ component even when on-shell, and one finds  for its couplings to fermions and pseudoscalars that $A/2 (1+ 4 z) +z =-1/2$ \cite{Ferrara:1976ni,PhysRevD.16.3427}.
In this work we will not concern ourselves with the details of the \enquote{off-shell parameters} because we are only interested in on-shell composite spin $3/2$ fermions; for the remainder of this work we set $\theta_{\mu\nu}=g_{\mu\nu}$ similar to Ref.~\cite{Garcia:2020hyo}.

Since  $\mathcal{B}$ fields caries  a Lorentz index and due to the constraint in Eq.~\eqref{eq:constr1} the higher dimensional operators destabilizing the dark baryon must involve derivatives. This is why the lowest allowed operator dimension for DM decay starts at $d=8$ compared to DM with spin   $1/2$, where e.g. dimension 7 operators are possible \cite{Garani:2021zrr}. Schematically the leading operators are at dark quark level
\begin{align}
    & \frac{c_8^{(1)}}{\Lambda_\text{UV}^4} \left(qqq\right)_\mu  \overline{L} \left( D^\mu \tilde{H}\right) \label{eq:eight} + \text{h.c.},\\
    & \frac{c_8^{(2)}}{\Lambda_\text{UV}^4} \left(qqq\right)_\mu   \left[\gamma^\alpha,\gamma^\beta\right]  \gamma^\mu  N_{L,R} B_{\alpha\beta}+ \text{h.c.},\label{eq:eight2}\\
    & \frac{c_{10}}{\Lambda_\text{UV}^6} \left(qqq\right)_\mu  N_{L,R} \left(\partial^\mu \overline{q}q\right) + \text{h.c.},\\
    & \frac{c_{11}}{\Lambda_\text{UV}^7} \left(qqq\right)_\mu  \overline{L}\left(\partial^\mu \overline{q}q\right)  \tilde{H} + \text{h.c.},
\end{align}
where $D^\mu$ denotes the gauge covariant derivative, $B^{\mu\nu}$ is the hypercharge field strength and $\left(qqq\right)_\mu$ indicates that we need a \enquote{spinor-vector} from the symmetric spin contraction of three dark quarks for spin $3/2$. 
The operators in Eq.~\eqref{eq:eight}-\eqref{eq:eight2} were already mentioned in Ref.~\cite{Dudas:2018npp} for the gravitino  and in Ref.~\cite{Garcia:2020hyo} for a general elementary spin $3/2$ fermion.

\twocolumngrid
\bibliographystyle{utphys}
\bibliography{references2}

\end{document}